\documentclass{raa}            

\usepackage{graphicx,times}             

\begin{document}

   \title{A first-order dynamical model of hierarchical triple stars and its application
 \,$^*$
 \footnotetext{$*$ This research is supported by the
National Natural Science Foundation of China under Grant Nos.
11178006 and 11203086.}
}

   \volnopage{Vol.0 (200x) No.0, 000--000}      
   \setcounter{page}{1}          

   \author{Xing-Bo Xu
      \inst{1,2}
   \and Fang Xia
      \inst{1}
   \and Yan-Ning Fu
      \inst{1}
   }

   \institute{ Purple Mountain Observatory, Chinese Academy of Sciences, Nanjing 210008, China;
              {\it xbxu@pmo.ac.cn; xf@pmo.ac.cn}\\
        \and   University of Chinese Academy of Sciences, Beijing 100049, China\\
   }

   \date{Received~~2015 month day; accepted~~2015~~month day}

\abstract{ For most hierarchical triple stars, the classical double two-body model of zeroth-order cannot describe the
  motions of the components under the current observational accuracy. In this paper, Marchal's first-order
  analytical solution is implemented and a more efficient simplified version is applied to real triple stars.
  The results show that, for most triple stars, the proposed first-order model
  is preferable to the zeroth-order model both
  in fitting observational data and in predicting component positions.
}

   \authorrunning{X.B. Xu, F. Xia \& Y.N. Fu }            
   \titlerunning{A first-order dynamical model of hierarchical triple stars}  

   \maketitle

%
%
\section{Introduction}           
\label{sect:intro}


  A hierarchical triple star is composed of a close binary
  and a distant third component.
  About one thousand stars of this kind are contained in the latest on-line version of
  \emph{The Multiple Star Catalog} (\cite{Tokovinin+1997}).
  In these systems, the primary components are usually bright.
  Bright stars are useful in many aspects (e.g. \cite{Urban+2014}).
  Though a set of isotropic and dense stars is crucial for some applications such as navigation,
  the stars with nearby companions are usually excluded. This is the case for the Hipparcos Celestial
  Reference Frame, as recommended in IAU resolution B1 (2000)\footnote{
  http://www.iau.org/static/resolutions/IAU2000\_French.pdf}.
  For triple stars,
  the problem lies mainly in that the primary positions generally cannot be predicted accurately
  by the almost exclusively used model, namely the classical double two-body model.

  Hierarchical triple stars are also of great interest in stellar physics
  and galactic astronomy, due to the fact that their dynamical evolution
  is important to both stellar and galactic evolutions
 (e.g. \cite{Binney+1998}, \cite{Valtonen+2006}, \cite{Aarseth+2003}).
 Moreover, these systems are often studied in terms of stability of the general three-body problem
 (e.g. \cite{MarchalBozis+1982}, \cite{LiFuSun+2009} ).
In some case studies, the results are sensitive to the mass parameters and the initial conditions
(e.g. \cite{OrlovZhuchkov+2005}),
the accuracies of which are limited again by the double two-body model
used in fitting observations (e.g. \cite{LiuRenXia+2009}) .

As a zeroth-order solution of the hierarchical three-body problem,
the double two-body model has the advantage of being analytical and simple.
The existing first-order analytical solutions are more accurate. The former one is still dominantly used,
while the latter ones, as far as we know, remain little used in fitting observations.
In this paper, the first-order solution by Marchal is efficiently
implemented. This is achieved mainly by making some simplified modifications
and high order approximations to Marchal's solution.
In the context of fitting observations of triple stars, we call Marchal's solution
and the double two-body solution, respectively, the M-model and the K-model.

   In section 2, the M-model is implemented.
   In section 3, the improvement in accuracy of M-model to K-model is statistically
   discussed with a set of sampling triple stars.
   In section 4, a simplified M-model is given and applied to real triple stars.
   Concluding remarks are given in the last section.

\section{An Implementation of M-model}
\label{sect:Model}

Consider a hierarchical three-body problem in an inertial coordinate system $\{\mathrm{O-}xyz\}$,
where $O$ is the center of mass and the $z$-axis parallel to the total angular momentum $\vec{C}$.
Denoting the masses of the inner two bodies by $m_{1}$ and $m_{2}$, and the mass of the third body by $m_{3}$,
we will use the following mass-dependent parameters,
\begin{eqnarray*}
   &  m_{t}=m_{1}+m_{2}+m_{3},  \quad m_{i}=\frac{m_{1}m_{2}}{m_{1}+m_{2}}, \quad
   m_{o}=\frac{(m_{1}+m_{2})m_{3}}{m_{t}},       \\
   &  \beta_{i}=\frac{G^{2}m_{1}^{3}m_{2}^{3}}{m_{1}+m_{2}}, \quad
   \beta_{o}=\frac{G^{2}(m_{1}+m_{2})^{3}m_{3}^{3}}{m_{t}},  \quad
   \beta_{1}=\frac{G^{2}(m_{1}+m_{2})^{7}m_{3}^{7}}{(m_{1}m_{2}m_{t})^{3}} \,,
\end{eqnarray*}
where $G$ is the gravitational constant.
 Let $\vec{r}$ be the position vector of $m_{2}$ relative to $m_{1}$,
and $\vec{R}$ the position vector of $m_{3}$ relative to the center of mass of the binary. The ratio $\varepsilon=\frac{r}{R}\equiv\frac{|\vec{r}|}{|\vec{R}|}$ is a small quantity.

The Delaunay variables as expressed in terms of the ordinary orbital elements $(a,e,i,\omega,\Omega,M)$ are
 \begin{eqnarray*}
 \begin{array}{llll}
 & \mathcal{L}_{i} =m_{i}\sqrt{G(m_{1}+m_{2})a_{i}} \,, & \mathcal{G}_{i} =\mathcal{L}_{i}\sqrt{1-e_{i}^{2}} \,,
 & \mathcal{H}_{i} =\mathcal{G}_{i}\cos i_{i}  \,,  \\
 &  \ell_{i} =M_{i} \,,          &  g_{i} =\omega_{i} \,,   &   h_{i} =\Omega_{i}   \,,  \\
 &  \mathcal{L}_{o} =m_{o}\sqrt{Gm_{t}a_{o}} \,,         &  \mathcal{G}_{o} =\mathcal{L}_{o}\sqrt{1-e_{o}^{2}} \,,
 &  \mathcal{H}_{o} =\mathcal{G}_{o}\cos i_{o} \,,  \\
 &  \ell_{o} =M_{o} \,,          &  g_{o} =\omega_{o} \,,   &   h_{o} =\Omega_{o} \,,
\end{array}
\end{eqnarray*}
where the subscripts $i$ and $o$ indicate the inner and outer orbits, respectively
 In these variables, the Hamiltonian up to the first order in $\varepsilon^2\sim(\frac{\mathcal{L}_{i}}{\mathcal{L}_{o}})^{4}$
 can be formally written as
 \begin{eqnarray}\label{Hami}
  H &=& H(\mathcal{L}_{i},\mathcal{G}_{i},\mathcal{L}_{o},\mathcal{G}_{o},\ell_{i},g_{i},
  \ell_{o},g_{o},\mathcal{H}_i+\mathcal{H}_o,h_o-h_i) \nonumber \\
   &\approx & H_{0i}+H_{0o}+H_{1} \nonumber \\
   & \equiv &  -\frac{\beta_{i}}{2\mathcal{L}_{i}^{2}} -\frac{\beta_{o}}{2 \mathcal{L}_{o}^{2}}
    +\frac{\beta_{1}}{2 \mathcal{L}_{o}^{2}}\frac{(1-e_{i}\cos E_{i})^{2}}{(1-e_{o}\cos E_{o})^{3}}\left(1-3\Phi^2 \right)\left(\frac{\mathcal{L}_{i}}{\mathcal{L}_{o}}\right)^{4} ,
 \end{eqnarray}
 where $\Phi=\Phi(\mathcal{L}_{i},\mathcal{G}_{i},\mathcal{L}_{o},\mathcal{G}_{o},\ell_{i},g_{i},\ell_{o},g_{o},\mathcal{H}_i+\mathcal{H}_o,h_o-h_i)
 =\frac{\vec{r}\cdot \vec{R}}{rR}$,
 $E_{i}=E_{i}(\mathcal{L}_{i},\mathcal{G}_{i};\ell_{i})$ and $E_{o}=E_{o}(\mathcal{L}_{o},\mathcal{G}_{o};\ell_{o})$
 are the eccentric anomalies of the inner and outer orbits, respectively.

In eq.(\ref{Hami}), $\mathcal{H}_o+\mathcal{H}_i$ and $h_{o}-h_{i}$ are understood as
two single canonical variables conjugating respectively to the negligible $h_{i}$ and $\mathcal{H}_{o}$.
And so, they are constants that can be calculated from the initial conditions.
The standard way to calculate the two negligible variables is by quadrature,
after all the other degrees of freedom are integrated. But in the present context,
we have as consequences of the integral of angular momentum
 \begin{eqnarray*}
  \mathcal{H}_i+\mathcal{H}_o=C \equiv |\vec{C}|, \quad  h_o-h_i=\pi, \quad
  \mathcal{H}_o=\frac{C^2+\mathcal{G}_{o}^{2}-\mathcal{G}_{i}^{2}}{2C} ,
\end{eqnarray*}
Therefore, only $h_i$ needs to be calculated by quadrature. Because of the short-period terms in the integrand,
the numerical quadrature is time-consuming. It is then preferable not to follow the standard way and
decouple only $(\mathcal{H}_o,h_o-h_i)$ from the other degrees of freedom at this stage.

For the system defined by the Hamiltonian eq.(\ref{Hami}) with $h_o-h_i=\pi$, a first-order integrable system
can be achieved by the Von Zeipel transformation (e.g. \cite{Harrington+1968}, 1969, \cite{CMarchal+1978}, 1990).
In the resulting canonical variables $(\mathcal{L}_{I},\mathcal{G}_{I},\mathcal{L}_{O},\mathcal{G}_{O},C; \ell_{I},g_{I},
  \ell_{O},g_{O},h_I)$, called long-period Delaunay variables, the new Hamiltonian 
  can be written as
\begin{eqnarray}\label{DoubleAvedH}
  \hat{H} &=& \hat{H}(\mathcal{L}_{I},\mathcal{G}_{I},\mathcal{L}_{O},\mathcal{G}_{O},C,g_{I})
     \nonumber  \\
  &=& \hat{H}_{0I}+\hat{H}_{0O}+\hat{H}_{1} \nonumber \\
    & \equiv &  -\frac{\beta_{i}}{2\mathcal{L}_{I}^{2}} -\frac{\beta_{o}}{2 \mathcal{L}_{O}^{2}}
    +\frac{\beta_{1}(3z-5)\mathcal{L}_{O}}{8\mathcal{G}_{O}^{3}} \left(\frac{\mathcal{L}_{I}}{\mathcal{L}_{O}}\right)^{4},
 \end{eqnarray}
  where
  \begin{eqnarray}\label{z}
   z=\frac{\mathcal{G}_{I}^{2}}{\mathcal{L}_{I}^{2}}
  \left[2-\biggl(\frac{C^2-\mathcal{G}_{I}^{2}-\mathcal{G}_{O}^{2}}{2\mathcal{G}_{I}\mathcal{G}_{O}}\biggr)^{2}\right]+
  5\left(1-\frac{\mathcal{G}_{I}^{2}}{\mathcal{L}_{I}^{2}}\right)
  \left[1-\biggl(\frac{C^2-\mathcal{G}_{I}^{2}-\mathcal{G}_{O}^{2}}{2\mathcal{G}_{I}\mathcal{G}_{O}}\biggr)^{2}\right]\sin^{2}g_{I} \,.
 \end{eqnarray}
 In this time-independent Hamiltonian of five degrees of freedom, there are four negligible variables $\ell_{I},\ell_{O},g_{O},h_{I}$.
 Their conjugate variables $\mathcal{L}_{I}$, $\mathcal{L}_{O}$, $\mathcal{G}_{O}$ and $C$,
 together with the total energy $\hat{H}$ and $z=z(\hat{H},\mathcal{L}_{I}, \mathcal{L}_{O}, \mathcal{G}_{O})$
 as given by solving eq.(\ref{DoubleAvedH}), are constants known from initial conditions.
 This confirms the integrability of the transformed Hamiltonian system.

 The differential equations for $\mathcal{G}_{I}$ and $g_I$, the variables of the only non-negligible degree of freedom,
 can be integrated simultaneously. But to be more efficient, we first integrate the equation for $\mathcal{G}_{I}$,
 decoupled from $g_I$ by using eq.(\ref{z}). In terms of $x=\frac{\mathcal{G}_{I}^{2}}{\mathcal{L}_{I}^{2}} \in (0,1)$,
 this equation writes
   \begin{eqnarray}\label{dxdtsign}
    \dot{x}=\pm\frac{3}{2}\frac{\beta_1 \mathcal{L}_{I}^4}{\mathcal{L}_{O}^3\mathcal{G}_{O}^3}\sqrt{P_{1}(x)P_{2}(x)},
   \end{eqnarray}
   where, with $A=\frac{C^{2}-\mathcal{G}_{O}^{2}}{2\mathcal{G}_{O}\mathcal{L}_{I}}$ and $B=\frac{\mathcal{L}_{I}}{2\mathcal{G}_{O}}$,
  \begin{eqnarray*}
  \begin{array}{l}
  P_{1}(x) =  B^{2}x^{2}-2(1+AB)x+z+A^2, \nonumber  \\
  P_{2}(x) = 4B^{2}x^{3}-(5B^{2}+8AB+3)x^{2}+(4A^{2}+10AB-z+5)x-5A^{2}.
  \end{array}
  \end{eqnarray*}
From the necessary condition $P_{1}(x)P_{2}(x)\geq 0$, Marchal (1990) pointed out that $x$ oscillates between two neighbouring roots,
$x_a \in (0,1)$ and $x_b \in (x_{a},1)$, of $P_{1}(x)P_{2}(x)$.
To be specific, the function $\dot{x}(t)$ defined in eq.(\ref{dxdtsign}) changes its sign from negative to positive at $x_a$,
and the opposite is true at $x_b$.

The difficulty in integrating eq.(\ref{dxdtsign}) caused by this
unfavorable feature of the right-hand side can be avoided. For this,
we introduce a continuously changing angular variable $\theta$, for
which $\mathrm{mod}(2 \pi)$ is not allowed, by the following
variable substitution $x=x_{a}+(x_{b}-x_{a})\sin^{2}\theta $.

 Let $\sigma_{3},\sigma_{4},\sigma_{5}$ be the other three roots of $P_{1}(x)P_{2}(x)$. We have
    \begin{eqnarray}\label{Tautheta}
     \frac{\mathrm{d}\tau}{ \mathrm{d}\theta}
     &=\mathcal{I}_{1}(\theta)\equiv \frac{1}{\sqrt{1-c_{1}\sin^{2}(\theta)+c_{2}\sin^{4}(\theta)-c_{3}\sin^{6}(\theta)}} \,,
    \end{eqnarray}
    where
    \begin{eqnarray*}
    \tau &=& \frac{3}{4}\frac{\beta_{1}\mathcal{L}_{I}^{4}}{\mathcal{L}_{O}^{3}\mathcal{G}_{O}^{4}}B \sigma \cdot  t \,,
    \quad \sigma=\sqrt{(\sigma_{3}-x_{a})(\sigma_{4}-x_{a})(\sigma_{5}-x_{a})}>0 \,,  \nonumber  \\
    c_{1} &=& d_{1}+d_{2}+d_{3}, \quad
    c_{2}=d_{1}d_{2}+d_{1}d_{3}+d_{2}d_{3}, \nonumber \\
    c_{3} &=& d_{1}d_{2}d_{3}>0, \quad d_{j}=\frac{x_{b}-x_{a}}{\sigma_{j+2}-x_{a}},(j=1,2,3).
    \end{eqnarray*}
    Given the initial condition $(t_{0},\theta_{0})$, the value of $\theta$ at any time
    $t$ can be obtained from an iterative method. And, given $\theta$, $\mathcal{G}_{I}(>0)$
    can be calculated from the defining formulae of $\theta$ and $x$.

 As $|\sin g_{I}(t)|$ can be solved from eq.(\ref{z}),
 the key to determining $g_{I}$ is its quadrant.
  Let $n$ be the biggest integer no greater than $2\theta/\pi$.
  The quadrant of $g_{I}(t)$
    can be deduced from the type of motion, $g_{I}(0)$ and $\theta$.
    Depending on the initial conditions, there are
    three types of motion.

  Type 1: $P_{2}(x_{a})=0$ and $P_{2}(x_{b})=0$. In this type of motion, $g_{I}$ oscillates
    around $\frac{\pi}{2}$ or $-\frac{\pi}{2}$ periodically.
    In the case of $\sin(g_{I}(0))>0$, $g_{I}(t)$ is in the first quadrant if $n$ is odd and the second quadrant if $n$ is even.
    In the other case, $g_{I}(t)$ is in the third quadrant if $n$ is odd and the fourth quadrant if $n$ is even.

  Type 2: $P_{2}(x_{a})=0$ and $P_{1}(x_{b})=0$. In this case, $g_{I}$ always increases as time grows.
  The $g_{I}(t)$ is in the same quadrant as
  $[\hat{\theta}_{n},\hat{\theta}_{n}+\frac{\pi}{2})$, where
  $\hat{\theta}_{n}=\frac{(n-1)\pi}{2}$ if $g_{I}(0)$ is in the same quadrant as $[-\frac{\pi}{2},\frac{\pi}{2})$ , and
  $\hat{\theta}_{n}=\frac{(n+1)\pi}{2}$ if $g_{I}(0)$ is in the same quadrant as $[\frac{\pi}{2},\frac{3\pi}{2})$ .

   Type 3: $P_{1}(x_{a})=0$ and $P_{2}(x_{b})=0$. The $g_{I}$ always decreases as time goes by.
  The $g_{I}(t)$ is in the same quadrant as
  $(\hat{\theta}_{n}-\frac{\pi}{2},\hat{\theta}_{n}]$, where
  $\hat{\theta}_{n}=\left(1-\frac{n}{2}\right)\pi$ if $g_{I}(0)$ is in the same quadrant as $(0,\pi]$, and
  $\hat{\theta}_{n}=-\frac{n\pi}{2}$ if $g_{I}(0)$ is in the same quadrant as $(-\pi,0]$.

   The other four angular variables can be obtained by quadrature,
   \begin{eqnarray}
   \begin{array}{lll}
   \ell_{I}(t) &=\ell_{I}(0)+\frac{\beta_{i}}{\mathcal{L}_{I}^{3}}t+
   \int_{\theta_{0}}^{\theta} F_{1}(x(\vartheta))\mathcal{I}_{1}(\vartheta)\mathrm{d}\vartheta, \quad
    & \ell_{O}(t) =\ell_{O}(0)+\frac{\beta_{o}}{\mathcal{L}_{O}^{3}}t+
    \frac{3}{8}\frac{\beta_{1}\mathcal{L}_{I}^{4}}{\mathcal{L}_{O}^{4}\mathcal{G}_{O}^{3}}(5-3z)t \,,
    \nonumber \\
    g_{O}(t) &=g_{O}(0)+\int_{\theta_{0}}^{\theta}F_{2}(x(\vartheta))\mathcal{I}_{1}(\vartheta)\mathrm{d}\vartheta, \quad
    & h_{I}(t) =h_{I}(0)+\int_{\theta_{0}}^{\theta} F_{3}(x(\vartheta))\mathcal{I}_{1}(\vartheta)\mathrm{d}\vartheta \,,
    \end{array}
   \end{eqnarray}
 where
 \begin{eqnarray}
 \begin{array}{ll}
  F_{1}(x) &=\frac{1}{B^{2}\sigma}
  \left[(z-\frac{5}{3})+\frac{x(z-2)+(A-Bx)^{2}}{2(1-x)}\right],
  \nonumber  \\
  F_{2}(x) &=\frac{5-3z}{2B\sigma}+
    \frac{1}{2B^{2}\sigma}
    \frac{(z-x)(A-Bx)}{x-(A-Bx)^{2}}
    \left[ 1+2B(A-Bx)\right],
    \nonumber \\
    F_{3}(x) &=-\frac{1}{2B^{2}\sigma} \frac{C}{\mathcal{G}_{O}}
    \frac{(z-x)(A-Bx)}{x-(A-Bx)^{2}} \,.
    \end{array}
 \end{eqnarray}

 If the first-order long-period solution is gotten,
 one can make inverse transformations of the solution to the original coordinate system.


\section{Comparison between M-model and K-model}
\label{sect:Num}

In order to compare the accuracy of different models in calculating the observational quantities,
it is necessary to do a numerical experiment. For the time being, we are interested in
only the systems with negligible 2nd-order perturbations. Therefore we generated 1000 systems,
which satisfy $|H_{2}|/|H_{0i}+H_{0o}+H_{1}|<0.01$ in $[-100,100]$ years, and
$H_{2}$ is the second-order perturbation term in the Hamiltonian (\ref{Hami}).
 This time span is used because the practical cycle of a star catalog is usually less than
 one hundred years.
 As expected, for some of the generated systems, especially for the
 systems with large periods and high eccentricities of the outer orbits,
 the first-order averaged perturbations are too large.
 For such a case, M-model fails to be the first-order model.
   We just consider the samples that satisfy
  \begin{eqnarray}\label{macon}
  |H_{1}/H_{0i}|<0.1, \quad |H_{1}/H_{0o}|<0.5 \,,
  \end{eqnarray}
  during $[-P_{t},P_{t}]$ years, where $P_{t}\geq \max(100,P_{o})$, and
  $P_{o}$ represents the initial period of the outer orbit.
  Nearly $90$ samples are excluded by condition eq.(\ref{macon}).
   In addition, Delaunay elements are not effective in describing the orbits that are
   near circular, near parabolic or near the reference plane, and M-model is not suitable to be used
   in coplanar motion. If there are very small divisors,
   the implicit Zeipel transformations can not be solved by the iterative method.
   Another $\sim 40$ samples are excluded, and $870$ samples remain.
   The remnant samples are used to do a numerical experiment to check the accuracy of M-model
  compared with K-model.

\begin{figure}
  \centering
   \includegraphics[width=10cm,keepaspectratio=true]{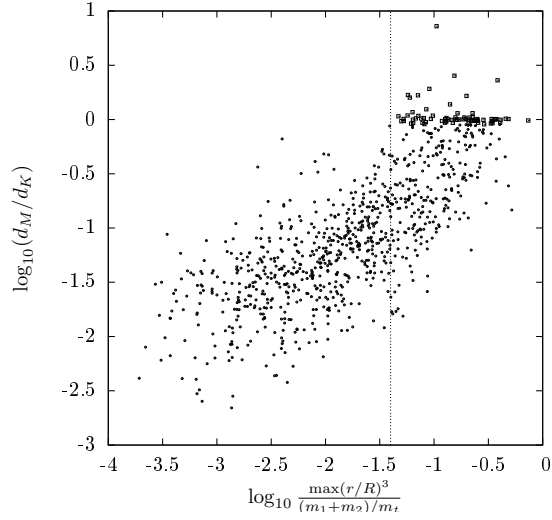}
  \caption{The abscissas on the x-axis are calculated in $[-P_{t},P_{t}]$ years. The abscissa of
  the dashed line is $-1.4$.
  Circular points represent the samples that satisfy $d_{M}/d_{K}<0.9$,
  while square points represent the samples that satisfy $d_{M}/d_{K}\geq 0.9$.
  There are $798$ circular points and $72$ square points. }\label{MS2271fig1}
  \end{figure}

  We calculate the positions of three bodies
  in the center-of-mass frame during the $[-100,100]$ years by M-model and K-model, respectively.
  As a comparison standard, these positions are also calculated by the numerical solution (\emph{N-model} for short).
  Denote the root-mean-squared errors (\emph{RMSE}) of the $9$-dimensional vectors of M-model
  relative to those of N-model by $d_{M}$, and the RMSE of the $9$-dimensional vectors of
  K-model relative to those of N-model by $d_{K}$.
  When $(r/R)^{3}\ll (m_{1}+m_{2})/m_{t}$, generally $d_{M}/d_{K}\ll 1$,
  as is shown in Fig. \ref{MS2271fig1}.
  Fig. \ref{MS2271fig1} shows that M-model is apparently
  better than K-model in accuracy when the abscissa is smaller than $-1.4$.
  When the abscissa is greater than $-1.4$, Fig. \ref{MS2271fig1}
  reveals that for most samples the M-model is still more accurate than K-model.

  For a few samples which are at the up-right quarter of Fig. \ref{MS2271fig1},
  the accuracy of M-model is not as good as that of K-model.
  The phenomena can be explained by the perturbations and the improper use
  of the Delaunay elements.

  There is one sample whose ordinate is apparently greater than $0.5$ in Fig. \ref{MS2271fig1}.
  We found that the outer orbit of this sample has a very large period and high eccentric.
  The $\max\frac{(r/R)^{3}}{(m_{1}+m_{2})/m_{t}}$ is really small during the considered $[-100,100]$ years, and
  K-model is very approximate to N-model. While M-model considers the averaged perturbations
  which are much greater. We calculated $\max|H-\hat{H}_{0i}-\hat{H}_{0o}-\hat{H}_{1}|$ in $[-P_{t},P_{t}]$ years and
  $\max|H-H_{0i}-H_{0o}|$ in $[-100,100]$ years. The former is more than $1000$ times of the latter, and this supports that
  M-model is not a first-order model in such cases.

  As the abscissas of samples represented by squared points are not sufficiently small (bigger than $-1.4$),
  the inaccuracies caused by small divisors cannot be ignored.
  For some samples represented by squared points in Fig. \ref{MS2271fig1},
  the detailed reasons are complex and uncertain currently.
  In all, M-model is better than K-model in accuracy for $\sim80\%$ of the samples, and
  can be credibly applied when the abscissa is smaller than $ -1.4$.


\section{The application}
\label{sect:application}

  Simplifications of M-model can be made according to the results of the numerical experiment.
  In eq.(\ref{Tautheta}), $x(\theta(t))$ can be solved efficiently by an approximation.
  Generally $\mathcal{I}_{1}(\vartheta)$ can be written
  \begin{eqnarray}
  \mathcal{I}_{1}(\vartheta)=\mathcal{I}_{2}(\vartheta)+
      \left[\mathcal{I}_{1}(\vartheta)-\mathcal{I}_{2}(\vartheta)\right] ,
  \end{eqnarray}
   where $\mathcal{I}_{2}(\vartheta)$ can be defined as
     \begin{eqnarray*}
      \mathcal{I}_{2}(\vartheta)=
      \left\{\begin{array}{ll}
      \frac{1}{\sqrt{1-c_{1}\sin^{2}\vartheta+c_{2}\sin^{4}\vartheta}} , &
        \textrm{if} \quad  c_{1}^{2}-4c_{2}>0, 1-c_{1}+c_{2} \gg c_{3}>0, c_{2}>0, \\
       \frac{1}{\sqrt{1-c_{1}\sin^{2}\vartheta}}, &
       \textrm{if} \quad c_{1}^{2}-4c_{2}\leq 0, 1-c_{1}\gg |c_{2}-c_{3}|>0,
       \end{array}\right.
     \end{eqnarray*}
     The formulas for calculating $\int_{0}^{\theta}\mathcal{I}_{2}(\vartheta)\mathrm{d}\vartheta$
     by elliptic functions can refer to \cite{ByrdFriedman+1971}. Similar studies which used elliptic functions can refer
     to \cite{Kozai+1962}, \cite{Soderhjelm+1982} and \cite{Solovaya+2003}.
     The remainder term $\mathcal{I}_{1}(\vartheta)-\mathcal{I}_{2}(\vartheta)$
     is generally small and sometimes can be ignored.
     If $\mathcal{I}_{1}(\vartheta)-\mathcal{I}_{2}(\vartheta)$ can be ignored,
      $\theta$ can be calculated analytically by elliptic functions.
     But here $\int_{\theta_{0}}^{\theta}
     \left[\mathcal{I}_{1}(\vartheta)-\mathcal{I}_{2}(\vartheta)\right]\mathrm{d}\vartheta$
     is considered by simple Newton-Cotes integration formula to make a better
     approximation.
     $\theta$ can be solved approximately by an iterative method.
     The three angular variables $\ell_{I},g_{O},h_{I}$
     can be integrated also by simple Newton-Cotes integration formula simultaneously.
     Another simplification is that the implicit Zeipel transformations from the averaged variables to the osculating elements
     can be turned into explicit. We call this model as \emph{MC-model}.

\begin{figure}
  \centering
   \includegraphics[width=13cm,keepaspectratio=true]{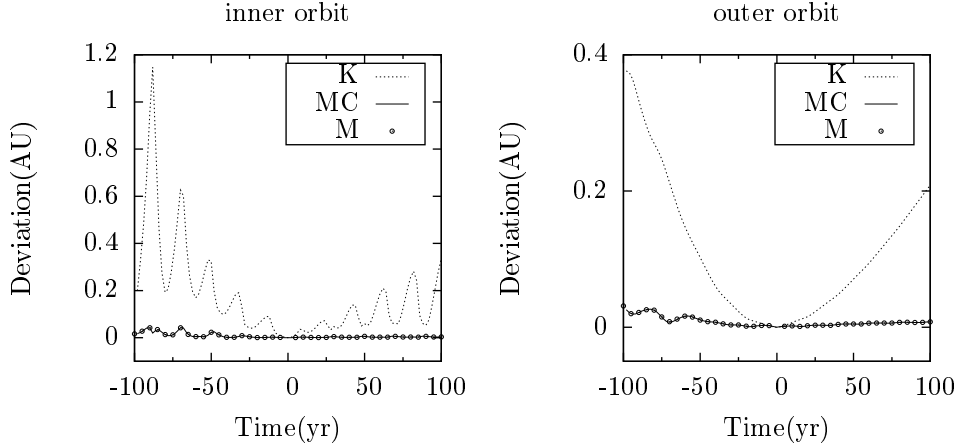}
  \caption{From the N-model, the deviations of component positions of WDS
02022+3643 calculated by using M-model, MC-model and K-model, respectively. }
\label{MS2271fig2}
  \end{figure}

We now apply this model to $25$ real triple stars with determined
dynamical state (component masses and kinematic parameters).  The
results are listed in Table \ref{result} including system name, order of magnitude of the perturbation
($\log_{10}\frac{\max[(r/R)^{3}]m_{t}}{m_{1}+m_{2}}$), the
RMSEs of M-model, K-model and MC-model, the  ratio of the RMSE of
MC-model to that of K-model ($\log_{10}\frac{d_{MC}}{d_{K}}$) and
the type of motion. According to this table, the accuracy between
M-model and MC-model is comparable.  For all these stars,  the RMSE
of MC-model in comparison with the K-model's, is reduced
significantly. Indeed, for $\sim60\%$ stars, the RMSEs are reduced by
more than one order of magnitude. To show more details, we take WDS
02022+3643 as an example. From the N-model, the deviations of
component positions calculated by M-model, MC-model and
K-model, respectively, are shown in Fig. \ref{MS2271fig2}. From this figure, we know
that the performance of MC-model is almost as good as M-model's.
 When compared with K-model, the model accuracy is significantly improved and the
applicable time span is significantly increased.

\begin{table}
\caption{The application results of the 25 observed triple stars
during the time span from 1900.0 to 2100.0.}\label{result}
\begin{tiny}
\begin{center}
\begin{tabular}{l c lll lc c c c c c }     
\hline\hline
system name      & perturbation order        & $d_{M}$       & $d_{MC}$     & $d_{K}$   &Improvement & Type                             \\
 (WDS) &  ($\log_{10}\frac{\max[(r/R)^{3}]m_{t}}{m_{1}+m_{2}}$)    & (AU)    & (AU)       &   (AU)     & ($\log_{10}\frac{d_{MC}}{d_{K}}$) &     (1/2)      \\
\hline
00325+6714                    & -1.52   & 2.77E-2  & 2.78E-2  & 0.47    & -1.23  & 2          \\
01148+6056                    & -4.64   & 6.11E-7  & 4.96E-4  & 1.96E-3 & -0.59  & 1          \\
02022+3643                    & -1.56   & 0.013    & 0.013    & 0.23    & -1.25  & 1          \\
03082+4057                    & -4.36   & 7.05E-4  & 1.43E-2  & 8.32E-2 & -0.76  & 2          \\
04142+2812                    & -4.13   & 4.78E-5  & 1.10E-4  & 0.10    & -2.96  & 1          \\
04400+5328                    & -1.53   & 0.119    & 0.119    & 0.96    & -0.91  & 2          \\
06262+1845                    & -7.69   & 2.13E-7  & 2.63E-6  & 6.01E-5 & -1.36  & 2          \\
07201+2159                    & -7.35   & 7.71E-9  & 7.58E-7  & 1.23E-5 & -1.21  & 2          \\
10373-4814                    & -2.88   & 2.60E-4  & 2.41E-3  & 2.72E-2 & -1.05  & 2          \\
10373-4814                    & -2.77   & 3.49E-4  & 4.73E-3  & 3.60E-2 & -0.88  & 2          \\
11308+4117                    & -6.22   & 1.23E-7  & 4.96E-6  & 4.06E-4 & -1.91  & 2          \\
12108+3953                    & -1.64   & 0.180    & 0.180    & 0.99    & -0.74  & 2          \\
12199-0040                    & -3.24   & 1.31E-3  & 3.26E-3  & 0.18    & -1.74  & 2          \\
15183+2650                    & -1.76   & 0.014    & 0.014    & 0.12    & -0.93  & 2          \\
16578+4722                    & -2.39   & 6.97E-4  & 8.63E-4  & 1.66E-3 & -0.28  & 2          \\
17539-3445                    & -7.14   & 4.58E-7  & 2.47E-5  & 9.87E-5 & -0.60  & 2          \\
19155-2515                    & -4.08   & 2.06E-5  & 2.03E-4  & 1.89E-2 & -1.97  & 1          \\
20396+0458                    & -1.45   & 7.17E-2  & 7.17E-2  & 1.30    & -1.26  & 1          \\
20475+3629                    & -2.15   & 1.19E-3  & 1.19E-3  & 5.27E-2 & -1.65  & 2          \\
22038+6437                    & -5.52   & 4.26E-7  & 4.91E-5  & 1.40E-4 & -0.46  & 2          \\
22288-0001                    & -4.03   & 2.95E-4  & 3.32E-4  & 2.44E-2 & -1.87  & 2          \\
22388+4419                    & -1.86   & 1.94E-2  & 1.94E-2  & 0.77    & -1.60  & 2          \\
23078+7523                    & -3.98   & 8.76E-6  & 1.18E-5 & 2.08E-3  & -2.25  & 2          \\
23393+4543                    & -1.77   & 5.05E-2  & 5.08E-2 & 0.72     & -1.15  & 2          \\
23393+4543                    & -1.86   & 5.53E-2  & 5.53E-2 & 0.45     & -0.91  & 2          \\
\hline
\end{tabular}
\end{center}
\end{tiny}
\end{table}

As we all know, one of the important factors decide the quality of
dynamical state determination is  the accuracy of the dynamical
model. In order to show the improvement in this respect brought by
the high accuracy MC-model, we apply both this model and K-model to
two systems, WDS 20396+0458 (HIP 101955, type 1) and WDS 00325+6714
 (HIP 2552, type 2).

 Two kinds of observations,  relative position data (RPD) and the Hipparcos
Intermediate Astrometric Data (HIAD) are used in the fitting. RPD
are extracted from the Washington Double Star (WDS) Catalog (\cite{mas01}),
and the Fourth Catalog of Interferometric Measurements of Binary
Stars (\cite{har01}).  HIAD are the abscissa residuals
with respect to a reference point, the abscissa of
which is calculated from a given solution. HIAD are read from the
resrec folder on the catalogue DVD
 of \cite{van07}. With these observational data, the maximum likelihood
 estimate of model parameters is obtained by minimizing the
objective function  ($\chi^{2}$)
\begin{eqnarray}\label{chi2}
\begin{array}{l}
\chi^{2}\equiv\sum_{i=1}^{N}(\frac{y_{i}-y(x_{i};a_{1}\cdots
a_{M})}{\sigma_{i}})^{2},
\end{array}
\end{eqnarray}
where $y_{i}$ is the observational quantity, $y(x_{i};a_{1}\cdots
a_{M})$ is the corresponding calculated value according to the
model parameters $a_{1}\cdots a_{M}$.  We use the Bounded Variable
Least Squares (BVLS) algorithm  (\cite{law95}) to minimize the $\chi^{2}$.

 HIP 101955 is a nearby low-mass triple star. There are
$15$ RPD points spanning from $1998$ to $2008$ of inner orbit, $46$ points
from $1934$ to $2008$ of the outer one, and $91$ HIAD in reference to  a
solution with $5$ parameters. In the previous determinations of the
dynamical state, the Kepler's two-body motion model is applied
separately to the inner $\{Aa,Ab\}$ and the outer $\{Am,B\}$ where
$Am$ is the center-of-mass of the inner binary AaAb (\cite{mal07}). The
results are collected in the \textit{Sixth Catalog of Orbits of
Visual Binary Stars}(ORB6) (\cite{har}), where the inner and outer orbits are
roughly evaluated as good and reliable, respectively, according to
the orbital coverage of the observations. Because more observations
are added, we firstly also use the K-model to fit observations. In
comparison with the previous results, the $\chi^{2}$ is found to be
reduced by $\sim66\%$. When the fitting model is replaced by MC-model,
the $\chi^{2}$ is further reduced by $\sim44\%$. Therefore, we conclude
that using high accuracy MC-model, the fitting result is
significantly better than the previous K-model's results. Using  the
fitted dynamical state parameters, the RMSEs of MC-model and
K-model are calculated during the forward $100$ years, that is, from
$2008$ to $2108$. The RMSE of MC-model in comparison with the K-model's,
is significantly reduced by more than $80\%$, from $35.9$mas (K-model)
 to $\sim 6.0$mas (MC-model). This result shows that though
starting with the same initial condition, for HIP 101955, the
K-model can not be used to predict the component positions.

For HIP 2552, there are $16$ RPD points spanning from $1989$ to $2005$ of
inner orbit, $75$ points from $1923$ to $2010$ of the outer one, and $151$
HIAD in reference to an acceleration solution with $7$ parameters. The
inner and outer orbits were provided by \cite{Doc08} and are evaluated
as good and indeterminate by ORB6. K-model is also firstly used to
fit the observations. In comparison with the previous fitting
results, the $\chi^{2}$ is reduced by $\sim42\%$.
When the fitting model is replaced by MC-model, though the $\chi^{2}$ is not
significantly reduced, the RMSE is reduced from $10.5$mas which is calculated by K-model
to $0.74$mas by MC-model. Using the fitted dynamical parameters, during the forward $100$
years, the RMSE of K-model is $29.8$mas while $\sim 5.0$mas
of MC-model. Therefore, K-model is also not suitable to predict the
component positions for HIP 2552.

\begin{figure}
  \centering
   \includegraphics[width=8cm]{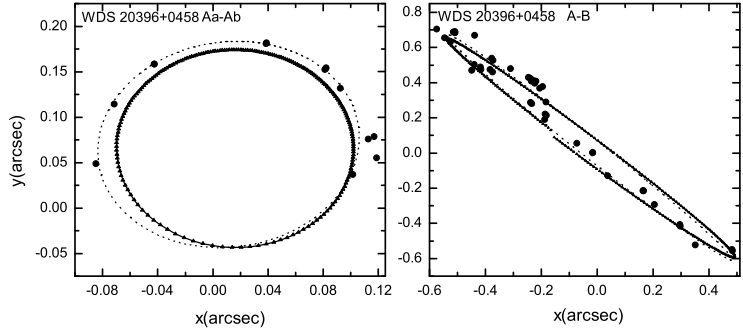}
  \caption{The fitting result of HIP 101955. }\label{MS2271fig3}
  \end{figure}

\begin{figure}
  \centering
   \includegraphics[width=8cm]{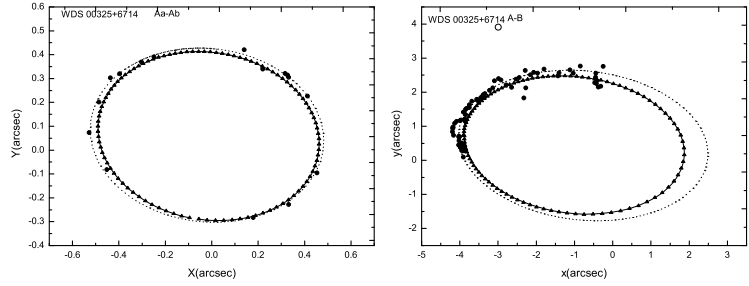}
  \caption{The fitting result of HIP 2552. The open circle is a discarded point.  }\label{MS2271fig4}
  \end{figure}

We plot the fitted trajectories of HIP 101955 and HIP 2552,
respectively, in Fig. \ref{MS2271fig3} and Fig. \ref{MS2271fig4}.
In these two figures, the filled circles are the RPD used in fitting, solid curves represent the
previous double two-body model while the dotted curves are the fitted trajectories calculated using the MC-model.
The trigonometric curves represent the N-model. As shown in the two figures, the difference between
MC-model and the N-model is small enough to be ignored.
The fitted dynamical state parameters and their 1$\sigma$ errors are listed in Table \ref{resultone}
and \ref{resulttwo}.

 \begin{table*}
\caption{The fitted dynamical masses and kinematic parameters of HIP
101955.}\label{resultone}
\begin{tiny}
\begin{center}
\begin{tabular}{c c c c c  }     
\hline\hline
parameter      &         &       &   & unit        \\
\hline
$\vec{M}$       & 0.786$\pm$0.11&0.493$\pm$0.11&0.516$\pm$0.21 &$M_{\odot}$ \\
$\vec{r}_{Ab}$      & -0.0598$\pm$0.0050&0.127$\pm$0.0050&-0.0188$\pm$0.022    &arcsec\\
$\vec{r}_{B}$&-0.188$\pm$0.0040&0.173$\pm$0.0038&0.902$\pm$0.025&arcsec\\
$\vec{v}_{Ab}$&-0.102$\pm$0.0062&-0.206$\pm$0.016&0.111$\pm$0.039 &arcsec/yr\\
 $\vec{v}_{B}$&0.0367$\pm$0.0027&-0.174$\pm$0.0066&0.0669$\pm$0.016&arcsec/yr\\
 \hline
\end{tabular}
\end{center}
\end{tiny}
\end{table*}

  \begin{table*}
\caption{The fitted dynamical masses and kinematic parameters of HIP
2552.}\label{resulttwo}
\begin{tiny}
\begin{center}
\begin{tabular}{c c c c c  }     
\hline\hline
parameter      &         &       &   & unit        \\
\hline
$\vec{M}$       & 0.389$\pm$0.038&0.0969$\pm$0.038&0.177$\pm$0.212 &$M_{\odot}$ \\
$\vec{r}_{Ab}$      & -0.0614$\pm$0.047&-0.298$\pm$0.029&0.290$\pm$0.032  &arcsec\\
$\vec{r}_{B}$&-4.029$\pm$0.016&0.609$\pm$0.015&-0.318$\pm$1.8&arcsec\\
$\vec{v}_{Ab}$&0.235$\pm$0.015&-0.0331$\pm$0.013&0.000668$\pm$0.025 &arcsec/yr\\
 $\vec{v}_{B}$&0.0478$\pm$0.0059&-0.0556$\pm$0.0029&0.0455$\pm$0.0097&arcsec/yr\\
 \hline
\end{tabular}
\end{center}
\end{tiny}
\end{table*}

\section{Conclusion and Discussion}

 Marchal's first-order analytical solution is implemented and a more efficient simplified version is
 applied to real hierarchical triple stars. The results show that the proposed first-order model is
 preferable to the classical double two-body model both in fitting observational data and
 in predicting component positions.

As pointed out in section \ref{sect:Num}, there are a few cases to which the M-model doesn't apply,
because of the inadequacy of the Delaunay elements. For these cases, Poincar\'{e} elements should be used instead.
There are also a few cases when the first-order perturbations are very small in the time span of observations,
but its maximum value over the whole period of the outer orbit is too large to apply M-model.
For these cases, our preliminary studies show that it is possible to give a suitable first-order solution
without resorting to averaging over the outer orbit.

\normalem
\begin{acknowledgements}
 The authors would like to thank the reviewers of this paper for their comments and suggestions,
 and also thanks to the editors of this journal. 
 This research is supported by the National Natural Science Foundation of China
 under Grant Nos. 11178006 and 11203086.

\end{acknowledgements}

\label{lastpage}

\end{document}